\documentclass[runningheads]{llncs}
\usepackage[caption=false]{subfig}
\usepackage{graphicx}
\usepackage{array}
\usepackage{multirow}
\usepackage[linesnumbered,ruled,vlined]{algorithm2e}
\begin{document}
\title{Predicting Outcome of Indian Premier League (IPL) Matches Using Machine Learning}
\titlerunning{Predicting Outcome of Indian Premier League (IPL) Matches}
%
\author{Rabindra Lamsal \and
Ayesha Choudhary}
\authorrunning{R. Lamsal and A. Choudhary}
%
\institute{School of Computer and Systems Sciences\\
 Jawaharlal Nehru University, New Delhi 110067\\
\email{rabindralamsal@outlook.com}, \email{ ayeshac@mail.jnu.ac.in}}
\maketitle              
\begin{abstract}
Cricket, especially the Twenty20 format, has maximum uncertainty, where a single over can completely change the momentum of the game. With millions of people following the Indian Premier League (IPL), developing a model for predicting the outcome of its matches is a real-world problem. A cricket match depends upon various factors, and in this work, the factors which significantly influence the outcome of a Twenty20 cricket match are identified. Each player’s performance in the field is considered to find out the overall weight (relative strength) of the team. A multivariate regression based solution is proposed to calculate points for each player in the league and the overall weight of a team is computed based on the past performance of the players who have appeared most for the team. Finally, a dataset is modeled based on the identified seven factors which influence the outcome of an IPL match. Six machine learning models were trained and used for predicting the outcome of each 2018 IPL match, 15 minutes before the gameplay, immediately after the toss. Three of the trained models were seen to be correctly predicting more than 40 matches, with Multilayer Perceptron outperforming all other models with an impressive accuracy of 71.66\%.

\end{abstract}
\section{Introduction}
With technology growing more and more advanced in the last few years, an in-depth acquisition of data has become relatively easy. As a result, Machine Learning is becoming quite a trend in sports analytics because of the availability of live as well as historical data \cite{analy1,analy2,analy3,analy4,analy5,analy6,analy7,analy8,analy9}. Sports analytics is the process of collecting past matches data and analyzing them to extract the essential knowledge out of it, with a hope that it facilitates in effective decision making. Decision making may be anything including which player to buy during an auction, which player to set on the field for tomorrow's match, or something more strategic task like, building the tactics for forthcoming matches based on players' previous performances.

Machine Learning can be used effectively over various occasions in sports, both on-the-field and off-the-field. When it is about on-the-field, machine learning applies to the analysis of a player’s fitness level, design of offensive tactics, or decide shot selection. It is also used in predicting the performance of a player or a team, or the outcome of a match. On the other hand, the off-the-field scenario concerns the business perspective of the sport \cite{sportb,sportb2}, which includes understanding sales pattern (tickets, merchandise) and assigning prices accordingly. The main focus is the healthy growth in business and profitability of the team owners and other stakeholders. On-the-field analytics generally make use of supervised machine learning algorithms, example: (i) regression for calculating the fitness of a player, (ii) classification for predicting an outcome of a match; while off-the-field analytics concerns around performing sentiment analysis to understand people’s opinion about a player or a team or a sport league. At present, Twitter has become one of the primary sources of data for sentiment analysis.

Sport Lisboa e Benfica, one of Portugal’s most successful football clubs advancing in the use of data modeling techniques while making decisions \cite{wired} is one real-world example of the use of machine learning in sports science. The club monitors and analyzes almost every aspect of a player, including their sleeping, eating, training habits. Once raw player data is recorded, various models are designed to analyze the data for optimizing match readiness and defining personalized practice schedules. With the application of machine learning and predictive analysis, the facts coming out of the devised models enable players to improve their performance continually. On the other cards, with those facts at hand, manager/coach gets a better idea about which player to be replaced, which player to be kept in the playing list and which player to be kept in the bench.

Major League Baseball (MLB) has seen enormous growth in the arena of sports analytics in the last few years \cite{baseballana1,baseballana2}. Professional MLB teams collect tremendous amount of ball-by-ball data and apply various machine learning approaches to get clear insights into the game, which is usually not visible through human analysis. Predicting the outcome of a match, classifying if a team will intentionally make a player walk at bat or classifying non-fastball pitches according to pitch type, etc. are some of the classification problems dealt using machine learning in baseball world \cite{baseballreview}.

Similarly, cricket has also been making use of sports analytics to perform prediction of outcome of a match, while the gameplay is in progress or before the match has even begun \cite{bandu,bailey,sankaranarayanan,cricai}. Even problem like predicting runs or wickets of a player for a match, based on his/her past performance is an interesting problem to work on. Some real-world tools which have been implemented in cricket include WASP (Winning and Score Predictor) \cite{wasp}, a tool which predicts a score and possible outcome of a limited over cricket match, i.e., One-day or Twenty20. Sky Sports New Zealand first introduced this tool in 2012 during an ongoing Twenty20 match. Technology like Hawk-Eye \cite{hawk1,hawk2,hawk3} which tracks the trajectory of a ball and visually displays the most statistically significant path, has also been officially in use as the Umpire Decision Review System since 2009. Similarly, other sports like tennis, badminton, snooker also make use of this computer-assisted intelligent technology.

\subsection{Machine Learning in Sports}
Various machine learning algorithms have been applied and tested for their efficiency in solving the problems in sports. The relation between machine learning and games dates back to the initial days of artificial intelligence when Arthur Samuel, a pioneer in the field of gaming and artificial intelligence studied machine learning approaches using the game of checkers \cite{samuel}.

A study \cite{hotspurs} was performed to predict the outcome (win, lose or draw) of football matches played by a professional English Premier League (EPL) team, Tottenham Hotspurs, based on matches that were played in the year between 1995 to 1997. It was observed that Bayesian networks relatively outperformed other machine learning algorithms which included MC4 - a decision tree learner, Naive Bayesian learner, Data-driven Bayesian, and K-nearest neighbor. The prediction accuracy of the Bayesian nets model was 59.21\%.

Match outcome prediction and game-play analysis are a prevalent problem that is tackled using machine learning. Another area where machine learning approaches are being used is extracting highlights from an on-going match. A study was performed to extract baseball match highlights on a set-top device \cite{rui}. The relative strength of classification algorithms, namely Support Vector Machine (SVM), Gaussian Fitting (GAU) and K-Nearest Neighbours (KNN) was considered for "excited speech" classification, and finally, SVM was applied. Six baseball matches covering 7 hours of game-play time was fed to the algorithm. 75\% of the highlights extracted by the algorithm were common with the highlights extracted manually by a human.

Just like in football, supervised machine learning algorithms have also been used in predicting the outcome of baseball matches. A project \cite{cs2291} used two learning methods, i.e., logistic classification and Artificial Neural Network (ANN) to predict the
result of the baseball post-season series. Although ANN came up with very poor accuracies, the accuracies out of the logistic model were satisfactory with training and test accuracies of 73.6\% and 62.6\% respectively. Another project applied four machine learning algorithms to understand career progression in Baseball \cite{cs2292}. The implemented algorithms were Linear Regression (Ridge Model), Multi-Layer Perceptron Regression (Neural Network), Random Forests Regression (Tree Bagging Model), Support Vector Regression (SVR). The dataset which was used to train these algorithms contained match data of the first six seasons of players' career. And the players' value were predicted. The prediction was near 60\% for the batters, while for pitchers the accuracy was very poor, i.e., something around 30-40\%.

\subsection{Machine Learning in Cricket}
In cricket, to predict an outcome of a match, the primary task is to extract out the essentials factors (features) which affect result of a match. Interesting works have been done in the field of predicting outcome in cricket. The literature survey concluded that the majority of the published works which predicted a result of a cricket match prior were for the test or one-day international cricket format.

Bandulasiri \cite{bandu} has analyzed the factors like home field advantage, winning the toss, game plan (first batting or first fielding) and the effect of Duckworth Lewis method \cite{dl} for one-day cricket format. Furthermore, Bailey and Clarke mention in their work \cite{bailey} that in one-day cricket format, home ground advantage, past performances, venue, performance against the specific opposition, current form are statistically significant in predicting total runs and predicting the outcome of a match. Similarly \cite{sankaranarayanan} discusses modeling home-runs and non-home runs prediction algorithms and considers taking runs, wickets, frequency of being all-out as historical features into their prediction model. But, they have not leveraged bowler’s features and have given more emphasis to batsmen. Kaluarachchi and Aparna \cite{cricai} have proposed a tool that predicts match apriori, but player performance has not been considered into their model.

\subsection{Indian Premier League}
\subsubsection{Introduction}
Indian Premier League (IPL) is a professional cricket league based on Twenty20 format and is governed by Board of Control for Cricket in India. The league happens every year with participating teams name representing various cities of India. There are many countries active in organizing Twenty20 cricket leagues. While most of the leagues are being overhyped and team franchises are routinely losing money, IPL has stood out as an exception \cite{iplgiant}. As reported by espncricinfo, with Star Sports spending \$2.5 billion for exclusive broadcasting rights, the latest season of IPL (2018, 11th) saw 29\% increment in the number of viewers including both the digital streaming media and television. The 10th season had 130 million people streaming the league through their digital devices and 410 million people watching directly on the TV \cite{livemint}. The numbers prove that IPL is a successful Twenty20 format based cricket league.

\subsubsection{Machine Learning in Indian Premier League}
Some interesting machine learning works have also been performed on data acquired from Indian Premier League matches. In a study \cite{saikia}, the Naive Bayesian classifier was used to classify the performance of all-rounder players (bowler plus batsman) into four various non-overlapping categories, viz., a performer, a batting all-rounder, a bowling all-rounder or an underperformer by being based on their strike rate and economy rate. Step-wise multinomial logistic regression (SMLR) was used to extract the essential predictors. When validated, the Naive Bayesian model was able to classify 66.7\% of the all-rounders correctly. The same authors later published a work in which an Artificial Neural Network model was used to predict the performance of bowlers based on their performance in the first three seasons of IPL \cite{saikia2}. When the predicted results were validated with actual performance of the players in season four, the developed ANN model had an accuracy of 71.43\%.

Although not related to IPL, a study performed at University College London in the area of predicting the outcome of a Twenty20 match \cite{county} would be a healthy addition as the literature work in the Twenty20 domain. The study made use of Naive Bayes, Logistic Regression, Random Forests, Gradient Boosting algorithms to predict the outcome of English County cricket matches. Two models were developed, each was given input of two different sets of features. The team only related features were input to the first model, while team and players related features were input to the second model. The study was concluded with Naive Bayes outperforming all other algorithms with the first model giving out average prediction accuracy of 62.4\% and second model giving average prediction accuracy of 63.9\%, i.e., 64\% average accuracy with 2009-2014 data and 63.8\% average accuracy with 2010-2014 data.

\subsection{Organization of Paper}
The paper is organized as follows: The proposed work is discussed in detail in various subsections of section 2. Results are shown in section 3, and section 4 concludes the paper.

\section{The Proposed Work}
The literature survey concluded that there was a need for a machine learning model which could predict the outcome of an IPL match before the game begins. Among all formats of cricket, Twenty20 format sees a lot of turnarounds in the momentum of the game. An over can completely change a game. Hence, predicting an outcome for a Twenty20 game is quite a challenging task. Besides, developing a prediction model for a league which is wholly based on auction is another hurdle. IPL matches cannot be predicted simply by making use of statistics over historical data solely. Because of players going under auctions, the players are bound to change their teams; which is why the ongoing performance of every player must be taken into consideration while developing a prediction model.

In sports, most of the prediction job is done using regression or classification tasks, both of which come under supervised learning. In simple terms, $y = f(x)$ is a prediction model which is learned by the learning algorithm from a set of dataset: $ D = {((X\textsubscript{1},y\textsubscript{1}), (X\textsubscript{2},y\textsubscript{2}), (X\textsubscript{3},y\textsubscript{3}), ... (X\textsubscript{n},y\textsubscript{n}))} $. Based on the type of output (y) supervised learning is divided further into two categories, viz., regression, and classification. In Regression, the output is a continuous value; however, classification deals with discrete kind of output. For predicting continuous values, Linear Regression appeared to be quite effective, and for classification problems like predicting the outcome of matches or classifying players, learning algorithms like Naive Bayes, Logistic Regression, Neural Networks, Random Forests were found being used in most of the previous studies.

In this work, the various factors that affect the outcome of a cricket match were analyzed, and it was observed that home team, away team, venue, toss winner, toss decision, home team weight, away team weight, influence the win probability of a team. The proposed prediction model makes use of multivariate Regression to calculate points of each player in the league and compute the overall strength of each team based on the past performance of the players who have appeared most for the team.

\subsection{The Prediction Model}
\subsubsection{Dataset}
The official website of Indian Premier League \cite{iplweb} was the primary source of data for this study. The data was scraped from the site and maintained in a Comma Separated Values (CSV) format. The initial dataset had many features including date, season, home team, away team, toss winner, man of the match, venue, umpires, referee, home team score, away team score, powerplay score, overs details when team reached milestone of multiple of 50 (i.e., 50 runs, 100 runs, 150runs), playing 11 players, winner and won by details. In a single season, a team has to play with other teams in two occasions, i.e., once as a home team and next time as an away team. For example, once KKR plays with CSK in its home stadium (Eden Gardens) next time they play against CSK in their home stadium (M Chinnaswamy Stadium). So, while making the dataset, the concept of home team and away team was considered to prevent the redundancy.

Indian Premier League has just been 11 years old, which is why only 634 matches data were available after the pre-processing. This number is considerably less with comparison to the data available relating to the test or ODI formats. Due to certain difficulties with some ongoing team franchises, in some seasons the league has seen the participation of new teams, and some teams have discontinued. Presence of those inactive teams in the dataset was not really necessary, but if the matches data were omitted where the inactive teams appeared, the chances were that the valuable knowledge about the teams which were still active in the league would deteriorate. For better understanding and to make the dataset look somehow cluttered-free, acronyms were used for the teams. Table \ref{table:acronym} lists the acronyms used in the dataset.

\begin{table}
\centering
\caption{Team names and their acronym.}\label{table:acronym}
\begin{tabular}{|p{5cm}|p{2cm}|}
\hline
\textbf{Team Name} & \textbf{Acronym}\\
\hline
Chennai Super Kings & CSK \\
\hline
Delhi Daredevils & DD \\
\hline
Kings XI Punjab & KXIP \\
\hline
Kolkata Knight Riders & KKR \\
\hline
Mumbai Indians & MI \\
\hline
Rajasthan Royals & RR \\
\hline
Royal Challenger Bangalore & RCB \\
\hline
Sunrisers Hyderabad & SRH \\
\hline
Rising Pune Supergiant & RPS$^a$ \\
\hline
Deccan Chargers & DC$^a$ \\
\hline
Pune Warriors India & PWI$^a$ \\
\hline
Gujrat Lions & GL$^a$ \\
\hline
Kochi Tuskers Kerala & KTK$^a$\\
\hline
\end{tabular}\\
\footnotesize{$^a$Inactive teams as of 2018 AD.}
\end{table}

\subsubsection{Calculating points of a Player}
There are various ways a player can be awarded points for their performance in the field. The official website of IPL has a \textit{Player Points} section where every player is awarded points based on these 6 features: (i) number of wickets taken, (ii) number of dot balls given, (iii) number of fours, (iv) number of sixes, (v) number of catches, and (vi) number of stumpings. To find out how IPL management was assigning points to each player based on these 6 features, a multivariate regression was used on the players' points data. Freedman \cite{stat} has beautifully explained the mathematics behind the Regression models. For this problem with six independent variables, the multivariate regression model takes the following form:

\begin{equation}
y = \beta\textsubscript{0} +  \beta\textsubscript{1}X\textsubscript{1} + \beta\textsubscript{2}X\textsubscript{2} + \beta\textsubscript{3}X\textsubscript{3} + \beta\textsubscript{4}X\textsubscript{4} + \beta\textsubscript{5}X\textsubscript{5} + \beta\textsubscript{6}X\textsubscript{6}
\label{equation:linear}
\end{equation}

Where, y is points awarded to a player, $\beta\textsubscript{0}$ is the bias term, $\beta\textsubscript{1}$ is the per wicket weight, $\beta\textsubscript{2}$ is the per dot ball weight, $\beta\textsubscript{3}$ is the per four weight, $\beta\textsubscript{4}$ is the per six weight, $\beta\textsubscript{5}$ is the per catch weight, and $\beta\textsubscript{6}$ is the per stumping weight.\\

When regression analysis was done on the player's point data, the following values were obtained for the weights $\beta\textsubscript{n}$ in equation \ref{equation:linear}:

\begin{center}
$\beta\textsubscript{0}$ = 0,  $\beta\textsubscript{1}$ = 3.5, $\beta\textsubscript{2}$ = 1, $\beta\textsubscript{3}$ = 2.5, $\beta\textsubscript{4}$ = 3.5, $\beta\textsubscript{5}$ = 2.5, $\beta\textsubscript{6}$ = 2.5
\end{center}

\subsubsection{Calculating Team weight}
For a team, there can be as many as 25 players. This is a limit put on by IPL governing council to the franchises. To find the average strength of a team, every player of the team is first sorted in the descending order according to their number of appearances in previous matches of the same season. Once players have been sorted, the top 11 players are considered for calculating the weight of the team because these players have played more games for the team and their performance influence the overall team strength.

\begin{equation}
weight\ of\ a\ team = \frac{\sum_{i = 1}^{11} i\textsuperscript{th} player's\ points}{total\ appearance\ of\ the\ team\ (ongoing\ season)}
\label{equation:weight}
\end{equation}

Now two more features, viz., home-team-weight and away-team-weight were also added to the previously designed dataset for all matches. Equation \ref{equation:weight} was used recursively to calculate the team weight based on the players who appeared the most for the team. Figuring team weight for all 634 matches was a tedious task. So, for example purpose, the final results of each season were considered, and the team weight for each team was calculated accordingly, and the same score was used for all the matches in that particular season. For better performance of the classifier, the team weight must be calculated immediately after the end of each match. This way, the real-time performance of each team and the newly computed weight can be used in predicting upcoming games.

\subsubsection{The final dataset}
In this study, Recursive Feature Elimination (RFE) algorithm was used as a feature selection method. As the name suggests, RFE recursively removes an unessential feature from a set of features, re-builds the model using the remaining features and recalculates the accuracy of the model. The process goes on for all the features in the dataset. Once completed, RFE comes up with top k number of features which influence the target variable (independent variable) at a level of extent. Sometimes, ranking the features and using the top k features for building a model might result in wrong conclusions \cite{rfe}. To prevent this from happening, the dataset was resampled, and RFE was operated in the subsets. The results were the same set of features obtained initially; hence, the initial set of features obtained from RFE did not seem to be biased. Using the RFE model, the number of features was reduced to 7. Thus obtained features which highly influenced the target variable were the home team, the away team, the venue, the toss winner, toss decision, and the respective teams' weight. Table \ref{table:finaldataset} shows the seven features considered for training the models.

\begin{table}
\centering
\caption{A portion of the dataset with the final set of features }\label{table:finaldataset}
\begin{tabular}{|p{1cm}|p{1cm}|p{1cm}|p{1cm}|p{5cm}|p{1cm}|p{1cm}|}
\hline
\textbf{h\_t$^a$} & \textbf{a\_t$^b$} & \textbf{t\_w$^c$} & \textbf{t\_d$^d$} & \textbf{stadium} & \textbf{w1$^e$} & \textbf{w2$^f$}\\
\hline
CSK &	RR &	RR & field & Dr DY Patil Sports Academy &101.75 &123.66\\
\hline
KKR &	DC &	DC	& bat&Eden Gardens&95.77&97.036\\
\hline
DD & KXIP & DD &  bat&Feroz Shah Kotla&105.93&114.93\\
\hline
RCB &	MI &	MI & field&M Chinnaswamy Stadium&90.89&102.21\\
\hline
\end{tabular}
\footnotesize{$^a$home team, $^b$away team, $^c$toss winner, $^d$toss decision, $^e$home-team-weight, $^f$away-team-weight}
\end{table}

\section{Results and Discussions}
A study carried out by Kohavi \cite{kohavi1995study} indicates that for model selection (selecting a good classifier from a set of classifiers), the best method is 10-fold stratified Cross-Validation (CV). This CV approach splits the whole dataset into k=10 equal partitions (folds) and uses a single fold as a testing set and union of other folds as a training set. The creation of folds is random. This process repeats for every fold. That means each fold will be testing set for once. Finally, the average accuracy is calculated out of the sample accuracy from each iteration.

Six commonly used classification-based machine learning algorithms \cite{witten}, viz., Naive Bayes\cite{langley}, Extreme Gradient Boosting \cite{xgboost}, Support Vector Machine \cite{svm}, Logistic Regression \cite{cox,walker}, Random Forests \cite{ho}, and Multilayer perceptron (MLP) \cite{haykin} were trained on the IPL dataset. The dataset contained all the match data since the beginning of Indian Premier League till 2017. The trained models were used to predict the outcome of each 2018 IPL match, 15 minutes before the gameplay, immediately after the toss. Table \ref{table:predictions} shows the performance of all classifiers. Among the six classification models, the MLP classifier outperformed all other classifiers by a notable margin in terms of prediction accuracy and weighted mean of precision-recall (F1 Score). The MLP classifier correctly predicted outcome of 43 matches of 2018 season, with classification accuracy of 71.66\% and F1 Score of 0.72. The precision, recall and F1 Score metrics for the MLP classifier is shown in Table \ref{table:fscore}.  Table \ref{table:hyperparameters} lists the hyper-parameters of the MLP classifier which were considered experimentally. Based on the classification accuracy, the MLP classifier was followed by Logistic Regression, Random Forests and SVM classifiers. However, Naive Bayes and Extreme Gradient Boosting classifiers performed poorly in predicting the outcomes of 2018 IPL matches.

\begin{table}
\centering
\caption{Performance of the classifiers on 2018 IPL matches }\label{table:predictions}
\begin{tabular}{|p{5cm}|p{3.5cm}|p{3cm}|}
\hline
\textbf{Classifier} & \textbf{Correct Predictions$^a$} & \textbf{Accuracy}\\
\hline
Naive Bayes & 30 matches & 50\%\\
\hline
Extreme Gradient Boosting & 33 matches & 55\%\\
\hline
Support Vector Machine & 38 matches&63.33\%\\
\hline
Logistic Regression & 41 matches&68.33\%\\
\hline
Random Forests & 41 matches&68.33\%\\
\hline
Multilayer perceptron & 43 matches&71.66\%\\
\hline
\end{tabular}
\footnotesize{$^a$out of 60 matches}
\end{table}

\begin{table}
\centering
\caption{Precision, Recall and F1 Score of the Multilayer perceptron classifier}\label{table:fscore}
\begin{tabular}{|p{1.5cm}|p{1.5cm}|p{1.5cm}|p{1.5cm}|}
\hline
 & \textbf{Precision} & \textbf{Recall} & \textbf{F1 Score}\\
 \hline
\textbf{0} & 0.62 & 0.59 & 0.60\\
\textbf{1} & 0.77 & 0.79 & 0.78\\
\textbf{average} & 0.71 & 0.72 & 0.72\\
\hline
\end{tabular}
\end{table}

\begin{table}[!h]
\centering
\caption{Hyper-parameters of the Multilayer perceptron}\label{table:hyperparameters}
\begin{tabular}{|p{6cm}|p{6cm}|}
\hline
\textbf{Hyper-parameter} & \textbf{number/value/function/type} \\
\hline
Number of hidden layers & 3\\
\hline
Number of hidden units in Layer 1/2/3 & 10 \\
\hline
Activation for hidden layers & ReLU function \\
\hline
Activation for the output layer & Sigmoid function \\
\hline
Optimizer & Adam \cite{adam}\\
\hline
Regularization & L2\\
\hline
Initial learning rate & 0.001\\
\hline
\end{tabular}
\end{table}

The MLP classifier was a three-hidden-layered artificial neural network with ten hidden units in each layer. The selection for the number of layers and the number of hidden units in each layer was made experimentally. The activation function in the hidden layer was Rectified Linear Unit (ReLU). Predicting the winner of a cricket match between a home team and an away team is a binary classification problem; hence, a sigmoid function was used as the activation function in the output layer.

\section{Conclusion}
In this study, the various factors that influence the outcome of an Indian Premier League matches were identified. The seven factors which significantly influence the result of an IPL match include the home team, the away team, the toss winner, toss decision, the stadium,  and the respective teams' weight. A multivariate regression based model was formulated to calculate the points earned by each player based on their past performances which include (i) number of wickets taken, (ii) number of dot balls given, (iii) number of fours hit, (iv) number of sixes hit, (v) number of catches, and (vi) number of stumpings. The points awarded to each player was used to compute the relative strength of each team. Various classification-based machine learning algorithms were trained on the IPL dataset designed for this study. The dataset contained all the match data since the beginning of Indian Premier League till 2017. The trained models were used to predict the outcome of each 2018 IPL match, 15 minutes before the game-play, immediately after the toss. The Multilayer perceptron classifier outperformed other classifiers by correctly predicting 43 out of 60, 2018 Indian Premier League matches.

The accuracy of the MLP classifier would have improved further if the team weight was calculated immediately after the end of each match. Because this is the only way, the classifier gets fed with real-time performance of the participating teams.

The Twenty20 format of cricket carries a lot of randomness, because a single over can completely change the ongoing pace of the game. Indian Premier League is still at infantry stage, it is just a decade old league and has way less number of matches compared to test and one-day international formats. Hence, designing a machine learning model for predicting the match outcome of an auction-based Twenty20 format premier league with an accuracy of 72.66\% and F1 score of 0.72 is highly satisfactory at this stage.

\section*{Acknowledgements} We want to show our gratefulness to Intel for providing us with a computing cluster for the period of this study.

\bibliographystyle{ieeetr}
\end{document}